\newcommand{\be}{\begin{equation}}\newcommand{\ee}{\end{equation}}
\newcommand{\bea}{\begin{eqnarray}}\newcommand{\eea}{\end{eqnarray}}
\newcommand{\nn}{\nonumber \\}\newcommand{\p}[1]{(\ref{#1})}

\newcounter{muni}
\newenvironment{remunerate}{\begin{list}{{\rm \arabic{muni}.}}
{\usecounter{muni}
\setlength{\leftmargin}{0pt}\setlength{\itemindent}{38pt}}}{\end{list}}
\documentstyle[amsfonts,12pt]{article}

\begin{document}
\begin{titlepage}
\begin{flushright}
LPTHE 98-44 \\
JINR E2-98-249 \\
hep-th/9810005 \\
October 1998
\end{flushright}
\vskip 1.0truecm
\begin{center}
{\large \bf Harmonic Space Construction of the Quaternionic\\
Taub-NUT metric}
\end{center}
\vskip 1.0truecm
\centerline{\bf Evgeny Ivanov${}^{(a)}$, Galliano Valent${}^{(b)}$}
\vskip 1.0truecm
\centerline{${}^{(a)}$\it Bogoliubov Laboratory of Theoretical 
Physics, JINR,}
\centerline{\it Dubna, 141 980 Moscow region, Russia}
\vskip5mm
\centerline{${}^{(b)}$ \it 
Laboratoire de Physique Th\'eorique et des Hautes Energies,}
\centerline{\it Unit\'e associ\'ee au CNRS URA 280,~Universit\'e Paris 7}
\centerline{\it 2 Place Jussieu, 75251 Paris Cedex 05, France}
\vskip 1.0truecm  \nopagebreak

\begin{abstract}
We present details of the harmonic space construction of a quaternionic 
extension of the four-dimensional Taub-NUT metric. As the main merit of 
the harmonic space approach, the metric is obtained in an explicit form 
following a generic set of rules. It exhibits 
$SU(2)\times U(1)$ isometry group and depends 
on two parameters, Taub-NUT `mass' and the cosmological constant. 
We consider several 
limiting cases of interest which correspond to special 
choices of the involved parameters.
\end{abstract} 
\end{titlepage}

\section{Introduction} 
One of the important implications of the harmonic (super)space 
method \cite{gikos} - \cite{gio} refers to the hyper-K\"ahler and 
quaternion-K\"ahler geometries where it provides an efficient way 
of explicit construction of the relevant metrics.

This approach was firstly introduced in the context of $N=2$ supersymmetry 
as the method of harmonic superspace \cite{gikos}. 
Its basic idea is to extend the standard $N=2$
superspace by a set of internal (`harmonic') variables 
$u^{\pm i}, u^{+i}u^-_i = 1$, parametrizing 
the coset space $S^2 \sim SU(2)/U(1)$ of the automorphism group $SU(2)$
of $N=2$ superalgebra. 
The main advantage of such a harmonic extension is the opportunity 
to single out a new subspace in it, the {\it analytic} subspace. 
It is closed under the supersymmetry transformations and contains half 
the Grassmann coordinates compared to the initial superspace. It was shown 
in \cite{gikos} 
that all $N=2$ theories admit an off-shell description  in terms of 
unconstrained harmonic analytic superfields having a clear geometric 
interpretation. 

Later on, it was realized that the harmonics can also be helpful in solving 
some purely bosonic geometric problems. 
It was shown in \cite{HK} that the constraints of the hyper-K\"ahler (HK) 
geometry (the vanishing of 
certain components of the curvature tensor) can be given an interpretation 
of the 
integrability conditions for the existence of analytic fields in a $SU(2)$ 
harmonic extension of 
the original $4n$-dimensional HK manifold $\{x^{i\mu}\}, 
(i=1,2; \mu = 1, ... 2n) $.  
This time, the $SU(2)$ to   
be `harmonized' is an extra $SU(2)$ rotating three complex structures of 
the HK 
manifold. A proper change of variables makes the half of covariant 
derivatives (their $u^+$  projections)  `short' that means the existence of 
an analytic subspace in the $SU(2)$-extended HK manifold. It is 
parametrized 
by $2n$ coordinates $x^{+\mu}$ and the harmonics $u^{\pm i}$. In the 
new basis the HK constraints can be solved in terms of unconstrained 
HK potential ${\cal L}^{+4}(x^{+\mu}, u^{\pm i})$. It encodes 
(at least, locally) all the information about the associated HK metric. 
It should be stressed that it allows one to {\it explicitly} compute 
the HK metric using a set of rules given in \cite{HK}. Indeed,using these 
techniques new derivations of the Taub-NUT metric 
\cite{cmp}, of the Eguchi-Hanson metric \cite{giot} and the full 
multicentre metric \cite{gorv} were given.

In \cite{gio}, a generalization of this approach to the important class of 
quaternion-K\"ahler (QK) 
manifolds was given. These manifolds generalize the HK ones in such 
a way that the extra $SU(2)$ group transforming complex structures 
becomes an 
essential part of the holonomy group. The HK manifolds can be treated 
as a degenerate case of the QK ones, with the $SU(2)$ part of the 
curvature 
tensor vanishing. One of important implications of QK manifolds is in 
General Relativity where four-dimensional QK manifolds appear as 
solutions of Einstein equation with non-zero cosmological constant. 
Besides, the QK manifolds appear as the target spaces of general $N=2$ 
supersymmetric sigma models coupled to $N=2$ supergravity \cite{bw}, 
with the $SU(2)$ curvature proportional to the Einstein constant. 

It was shown in \cite{gio}, that, using the concept of harmonic extensions, 
the QK geometry constraints can be solved quite similar to those of the HK 
geometry. 
Once again, the solution is given  
in terms of
some unconstrained potential ${\cal L}^{+4}$ living on the analytic 
subspace 
parametrized by $SU(2)$ harmonics and half of the original space 
coordinates. 
The specificity of the QK case manifests itself in the presence of 
non-zero 
constant 
$SU(2)$ curvature on all steps  of the road from  ${\cal L}^{+4}$ to the 
related 
QK metric. This somewhat obscures the computations, though they remain more
 or 
less direct. It is interesting to consider some examples in order to see 
in detail how the 
machinery proposed in \cite{gio} works. Only the simplest case of the 
homogeneous QK manifold  $Sp(n+1)/Sp(1)\times Sp(n)$ (corresponding to the 
flat HK manifold) was considered as an example in ref. \cite{gio}. 

The aim of this paper \footnote{It is an extended detailed version of 
ref. \cite{iv1}.} is to demonstrate the efficiency of the harmonic 
space approach on the less trivial example of a quaternionic 
generalization of the four-dimensional Taub-NUT metric (QTN in 
what follows). In section 2 we give a brief summary of the concepts and 
relations 
of ref. \cite{gio} adapted to our practical purpose. Then in sections  
3 and 4 we present 
the actual computations. Like in the HK case \cite{cmp}, they are greatly 
simplified due to the `translational' $U(1)$ isometry of the QTN metric 
(its full isometry, like in the 
HK case, is $SU(2)\times U(1)$, with the $SU(2)$ factor being `rotational' 
isometry). 
The metric obtained depends on two parameters, the 
TN `mass' parameter and the constant $SU(2)$ curvature parameter which 
can be interpreted as the inverse `radius' of the corresponding `flat' QK 
background $\sim Sp(2)/Sp(1)\times Sp(1)$. In section 5 we perform the 
identification of the metric obtained from harmonic space with 
its standard form given in the literature \cite{egh} and 
consider some important limits related to special choices of the 
parameters. 
         
\setcounter{equation}{0}
\section{Generalities}
Here we sketch the basic points of the construction of 
\cite{gio} (with minor deviations in the notation). We refer the reader 
to \cite{gio} for the detailed explanations and proofs. 

One starts with a $4n$-dimensional Riemann manifold with local coordinates 
$\{x^{\mu m}\}, \mu = 1,2, ..., 2n; m = 1,2$ and uses a vielbein formalism. 
The QK geometry can be defined as a restriction of the general Riemannian 
geometry in $4n$-dimensions, such that the holonomy group of the latter is 
required to be a subgroup of $Sp(1)\times Sp(n)$\footnote{For 
$4$-dimensional case this definition ceases to be meaningful, and it is 
replaced by the requirement that the totally symmetric part of the $Sp(1)$ 
component of the curvature tensor lifted to the tangent space is 
vanishing.}. 
Thus one can choose the tangent space group from the very beginning to be 
$Sp(1)\times Sp(n)$ and define the QK geometry via appropriate 
restrictions on the curvature tensor lifted to the tangent space 
(taking into account that the holonomy group is generated 
by this tensor). As explained in \cite{gio},  
for the QK manifold of any dimension the defining constraints can 
be concisely written as a restriction on the form of the commutator
of two covariant derivatives
\be 
\left[ {\cal D}_{\alpha (i}, {\cal D}_{\beta k)} \right] =
 -2 \Omega_{\alpha\beta}
R \Gamma_{(ik)}~. \label{constr}
\ee         
Here 
\be \label{covD}
{\cal D}_{\alpha i}  = e^{\mu m}_{\alpha i}(x) \nabla_{\mu m}  
= e^{\mu m}_{\alpha i}(x)\;\frac{\partial}{\partial x^{\mu m}} 
+ [Sp(1)\times Sp(n)-\mbox{connections}] ~,
\ee
$e^{\mu m}_{\alpha i}(x)$ being the relevant 
$4n \times 4n$ vielbein with the indices $\alpha = 1,2, ... 2n$ and $i=1,2$ 
rotated, respectively, by the local tangent $Sp(n)$ and $Sp(1)$ groups, 
$\Omega_{\alpha\beta}$ is the $Sp(n)$-invariant skew-symmetric tensor 
serving to raise and lower the $Sp(n)$ indices 
($\Omega_{\alpha\beta}\Omega^{\beta\gamma} = \delta^\gamma_\alpha $), 
$\Gamma_{(ik)}$ are the $Sp(1)$ generators, and  $R$ is a constant, 
remnant of the $Sp(1)$ component of the Riemann tensor (its constancy is 
a consequence of the QH geometry constraint and Bianchi identities). 
The scalar curvature coincides with $R$ up to a positive 
numerical coefficient, so the cases $R>0$ and $R<0$ correspond to compact 
and non-compact manifolds, respectively. 
In the limit $R=0$ eq. \p{constr} is reduced to the constraint defining 
the HK geometry \cite{HK}, in accord with the interpretation of HK manifolds 
as a degenerate subclass of the QK ones. 
 
Like in the HK case \cite{HK}, in order to explicitly figure out what 
kind of restrictions is imposed by (1) on the vielbein 
$e^{\mu m}_{\alpha i}(x)$ and, hence, on the metric 
\be  \label{metr}
g^{\mu s \; \nu m} = e^{\mu s}_{\alpha i}\; e^{\nu m \;\alpha i}~, \quad 
g_{\mu s \; \nu m} = e_{\mu s \;\alpha i}\; e_{\nu m}^{\alpha i}~,
\ee
one should solve the constraints \p{constr} 
by regarding them as an integrability condition along some complex 
directions in a harmonic extension of the original manifold. Due to 
the non-vanishing r.h.s. in (1), the road to such an interpretation 
in the QK case is more tricky and passes through two steps. 

First, one introduces a sort of harmonic variables $v^a_i$, 
$(v^a_iv^{b\;i} = \epsilon^{ab})$, on the local $Sp(1) \sim SU(2)$ 
group acting in the tangent space and represents the generators 
$\Gamma_{(ik)}$ as differential operators in these extra variables. 
Correspondingly, the $Sp(1)$-parts of the gauge connection acquires 
the meaning of extra components of vielbein, while the $Sp(1)$-part 
of the curvature tensor becomes the relevant torsion (of course, no 
torsion appear in the $x$-directions). 
At this step, one converts all the tangent $Sp(1)$ indices $i,k,..$ 
with the $Sp(1)$ harmonics and so replaces them by the indices 
$a,b,c..$ on which some rigid $SU(2)$ is realized (they are analogs of 
the indices $\pm $ of the standard $SU(2)$ harmonics, while the rigid 
$SU(2)$ is an analog of the right $SU(2)$ acting on these $\pm $). 

Next, one introduces the second set of harmonics on the rigid 
$SU(2)$, that time the standard $S^2 \sim SU(2)/U(1)$ ones 
$u^{\pm a}, u^{+ a}u_a^- = 1$.  
The parametrization of the bi-harmonic space 
$Sp(1)\times S^2 = \{v_i^a, u^{\pm a}\}$ which is most convenient for 
our purposes is as follows \cite{gio}
\be \label{Zparam}
Sp(1)\times S^2 = \{ z^0, z^{++}, z^{--}, w^+_i = u^+_i - z^{++}u^-_i, 
w^-_i = u^-_i \}\;.
\ee
One can define three independent covariant derivatives $Z^{++}, Z^{--}, 
Z^0$ 
on the $Sp(1)$ part of the bi-harmonic space and three standard harmonic 
derivatives $D^{++}, D^{--}, D^0 $ on its $SU(2)$ part. They 
form a semi-direct sum of 
two $su(2)$ algebras. The operator $D^0$ is the customary harmonic 
$U(1)$-charge counting operator, while the operator $Z^0 = 
z^0\partial/\partial z^0$ commuting with $D^0$ was called in 
\cite{gio} the `$Sp(1)$ weight operator': it counts the degree of 
homogeneity 
in the $Sp(1)$ coordinate $z^0$ (this coordinate is assumed 
to be $1$ at the origin of the $Sp(1)$ group manifold, so, 
in a vicinity of the origin, one can consider both its positive 
and negative powers).

Successively projecting the constraint (1) on the first and second sets 
of harmonics, one puts it into the following equivalent 
suggestive form \cite{gio}
\be  \label{constr2}
\left[{\cal D}_{\alpha}^+, {\cal D}_{\beta}^+ \right] = 
- 2\Omega_{\alpha\beta}R Z^{++}~. 
\ee   
In the basis \p{Zparam} 
\be 
{\cal D}_{\alpha}^+  = z^0 w^{+i} {\cal D}_{\alpha i} = 
z^0 w^{+i}e^{\mu m}_{\alpha i}(x)\partial_{\mu m} +
\dots  \equiv z^0 E^{+\mu m}\partial_{\mu m} + ...~, \quad 
Z^{++} = (z^0)^2 \frac{\partial}{\partial z^{--}}~,  \label{Dstruct}
\ee
where `dots' stand for the $Sp(n)$ connection and new 
$Sp(1)$ vielbeins terms.
Eq. \p{constr2}, together with the evident relation 
\be \label{ZDcomm}
[Z^{++}, {\cal D}_{\alpha}^+] = 0~, 
\ee
form a closed algebra of integrability conditions implying the existence 
of {\it analytic} fields on the bi-harmonic 
extension $\{x, z, w\}$ of the QK manifold $\{x \}$
\bea 
&& {\cal D}_{\alpha}^+ \Phi^{(q,p)}(x,z,w) = 0~, \label{anal} \\
&& \frac{\partial}{\partial z^{--}} \Phi^{(q,p)}(x,z,w) = 0~. \label{anal1}
\eea
Such fields can have a definite $U(1)$ charge $q$ and  
$Sp(1)$ weight $p$ 
\be  \label{qp}
D^0 \Phi^{(q,p)} = q \Phi^{(q,p)}, \quad Z^0 \Phi^{(q,p)} = p 
\Phi^{(q,p)} \;\; \Rightarrow \;\;\Phi^{(q,p)} = (z^0)^p \Phi^q
\ee
due to the commutation relations 
\bea 
[D^0, {\cal D}_{\alpha}^+] &=& [Z^0, {\cal D}_{\alpha}^+] = 
{\cal D}_{\alpha}^+~, \;
[D^0, Z^{++}] = 2 Z^{++}~, \; [Z^0, Z^{++}] = 2 Z^{++}~, \nn  
\left[D^0, Z^0\right] &=& 0~. \label{comm}
\eea

Eqs. \p{anal}, \p{anal1} mean that $ \Phi^{(q,p)}(x,z,w)$ effectively 
depend only 
on half of the $x$ variables and do not depend on $z^{--}$. One can 
covariantly eliminate the dependence on $z^{++}$ as well, 
taking into account the relations \cite{gio}
\bea
&& D^{--} - Z^{--} = \frac{\partial}{\partial z^{++}}~, \label{def1a} \\ 
&& [D^{--} - Z^{--}, {\cal D}_{\alpha}^+] = 
[D^{--} - Z^{--}, Z^{++}] = 0~. \label{comm2} 
\eea
Due to these relations (and appropriate ones involving $D^0$ and $Z^0$), 
the 
additional analyticity condition 
\be
(D^{--} - Z^{--}) \Phi^{(q,p)} = 
\frac{\partial}{\partial z^{++}} \Phi^{(q,p)} = 0 \quad 
\Rightarrow  \Phi^{(q,p)} (x,z,w) = (z^0)^p \Phi^q (x,w)~, 
\label{anal2}
\ee
is compatible with eqs. \p{anal}, \p{anal1}, \p{qp}. 

Thus, the original constraints \p{constr} amount to the existence 
of some $z$-independent invariant subspace $\{x, w\}$ in the bi-harmonic 
extension $\{ x, v, u \} = \{x, z, w \}$ of the original QK manifold 
$\{ x \}$. Moreover, as was already mentioned, eqs. \p{constr2}, \p{anal} 
imply an opportunity to extract 
a smaller subspace $\{x^{+\mu}_A, w^{\pm i}_A \}$ containing only half 
the original $x$-coordinates. 
Indeed, the covariant derivatives which introduce the QK analyticity 
via eqs. \p{anal}, \p{anal1}, \p{qp} and \p{anal2}, 
\be \label{CR}
{\cal D}_{\alpha}^+~, \quad Z^{++} = 
(z^0)^2\frac{\partial}{\partial z^{--}}~,  \quad  
D^{--} - Z^{--} = \frac{\partial}{\partial z^{++}}~, \quad 
Z^0 = z^0\frac{\partial}{\partial z^{0}}~,
\ee
form a closed algebra. So, by the Frobenius theorem, there 
should exist a basis in the space $\{x, z, w \}$, such that these 
derivatives 
are reduced to the partial ones in this basis, i.e. become `short'. 
The $Sp(1)$ derivatives are already short, so it remains to make short 
the derivative ${\cal D}_{\alpha}^+$. This can be done by the following 
change of variables \cite{gio} (we do not give the change of $z^{--}$ as 
irrelevant for our purposes)
\bea 
&& z^{++}_A = z^{++} + v^{++}(x,w)~, \quad z^0_A = t(x,w)z^0~, 
\label{zbridge} \\
&& w^{+\;i}_A = w^{+\;i} -v^{++}(x,w)w^{-\;i}~, \quad w^{-\;i}_A = 
w^{-\;i}~, 
\label{wbridge} \\
&& x^{\pm \mu}_A = x^{\mu\; n}w^{\pm}_n + v^{\pm \mu}(x,w)~. 
\label{xbridge}
\eea
Here, $v^{++}(x,w)$, $t(x,w)$ and $v^{\pm \mu}(x,w)$ are `bridges' from 
the `$\tau $- world' where the QK analyticity is implicit, to the 
`$\lambda $-world' where it is manifest. The bridges can be 
consistently chosen independent of $z$-coordinates, once again due to 
the above integrability relations. One should accompany \p{zbridge} - 
\p{xbridge} by a rotation of all $Sp(n)$ indices with an 
appropriate matrix bridge removing the $Sp(n)$ connection part from 
${\cal D}_{\alpha}^+$. The existence of such a bridge is also guaranteed 
by the aforementioned Frobenius theorem. As a result,  
${\cal D}_{\alpha}^+$ in the $\lambda $-world is given by 
\be  \label{Danal}
({\cal D}_{\alpha}^+)_\lambda = z^0_A \left(E^\mu_\alpha (x,w)
\frac{\partial}{\partial x^{-\mu}_A} + \rho^{-}_{\alpha}(x,w) 
\frac{\partial}{\partial z^{--}_A}\right)~, 
\ee 
where the vielbeins are still functions over the whole space 
$\{ x, w \} = \{x_A, w_A \}$. In what follows, we  will ignore the terms 
containing $z^{\pm\pm}$-derivatives, since one can restrict the 
consideration 
to $z$-independent functions in view of the $Sp(1)$ analyticity  
conditions \p{anal1}, \p{anal2} and their $\lambda $-world counterparts. 
As far as we are interested in the QK metric, we can also assume these 
functions to have $p=0$, i.e. ignore the $Sp(1)$ weight terms. 

Before going further, let us briefly explain the motivation of introducing 
extra $Sp(1)$ bridges $v^{++}(x,w)$, $t(x,w)$ in the QK case compared 
to the 
HK case where no such objects appear \cite{HK}. The point is that 
the $Sp(1)$ vielbein part of ${\cal D}_{\alpha}^+$ in the original basis 
contains a term $\sim \partial^{--}_W$ in parallel with 
the terms entering with $z$-derivatives \cite{gio}. The shift of $z^{++}$ 
and, respectively, of $w^{+}_i$ in \p{zbridge}, \p{wbridge} is necessary 
to ensure such term to be absent in the analytic basis. Simultaneously, 
the term $\sim \partial/\partial z^{++}$ is removed. Analogously, 
the bridge $t(x,w)$ is introduced in order to remove the terms $\sim Z^0$ 
from  ${\cal D}_{\alpha}^+$. 

We wish to stress that the introduction of 
$z$-coordinates is, to some extent, an auxiliary intermediate step. It 
is basically intended to consistently define the analyticity underlying 
the QK case and to arrive at the `true' harmonic variables $w^{+\;i}_A 
= u^{+ \;i} - z^{++}_A u^{-\;i} = 
w^{+ \;i} - v^{++}(x,w) u^{-\;i}$. The only manifestation of $z$ 
coordinates is the presence of non-trivial harmonic part 
in the analyticity-preserving diffeomorphism group of 
$\{x^{+\mu}_A, w^{\pm\;i}\}_A$, $\delta w^{+\;i}_A = 
\lambda^{++}(x^+_A, w_A)\;w^{-\;i}_A$, 
along with the analytic diffeomorphisms $\delta x^{+\mu}_A = 
\lambda^{+\mu}_A(x^+_A, w_A)$. The basic relations of the QK geometry 
in the $\lambda $-world, including the formula for the QK metric, 
contain no any trace of these auxiliary coordinates \cite{gio}. 

Next steps towards solving the QK constraints mimic those in 
the HK case \cite{HK} \footnote{Quite analogous reasoning is employed 
while solving the constraints of self-dual Yang-Mills theory in 
the harmonic space approach \cite{HK1} and those of $N=2$ supersymmetric 
Yang-Mills theory \cite{gikos}.}. The basic incentive is to show that the 
bridges, as well as the $\lambda $-world vielbein, can be expressed in 
terms of some unconstrained analytic object, the QK potential. 
Afterwards, the $\tau $-world vielbein and, hence, the sought QK metric 
can be restored from the $\lambda $-world ones by the change of variables 
inverse to \p{xbridge}. 
 
First, one observes that the particular harmonic 
dependence of the $\tau $-world ${\cal D}_{\alpha}^+$, \p{Dstruct}, 
amounts to imposing the relation 
\be  \label{DDcom}
[D^{++}, {\cal D}_{\alpha}^+] = 0~,
\ee
along with eqs. \p{ZDcomm}, \p{comm}, \p{comm2}. In the $\tau $-basis 
$D^{++}$ contains no derivatives with respect to $x$-coordinates, being 
simply $D^{++} = \partial^{++}_u $ in the parametrization $\{x, v, u\}$. 
After passing to the $\lambda $-basis by eqs. \p{zbridge} - \p{xbridge}, 
it acquires induced vielbeins (together with some induced gauge 
$Sp(n)$ connection as a consequence of the rotation by a matrix 
$Sp(n)$ bridge). Ignoring the $z_a$-terms and $Sp(n)$-connection, it 
reads  
\be
(D^{++})_\lambda = \partial^{++}_A + H^{+3\mu}\partial^-_\mu + 
x^{+\mu}\partial^+_\mu + H^{+4}\partial^{--}_A = 
\partial^{++}_w + v^{++} D^0~,  \label{D++lambd}
\ee
where $\partial^{\pm}_\mu \equiv \partial/\partial x^{\mp\mu}_A$ and 
$H^{++-\mu} = x^{+\mu}$ as a result of partial fixing of the 
$x^{-\mu} \break $ -diffeomorphisms gauge freedom \cite{gio} (in what 
follows, 
for brevity, we omit the subscript `A' where this cannot cause a confusion). 
The induced vielbeins are related to the bridges as follows 
\bea
&& (\partial^{++}_w + v^{++}) x^{+\mu}_A = H^{+3\mu}  \label{H3} \\
&& (\partial^{++}_w + v^{++}) v^{++} = - H^{+4}~. \label{H4}
\eea
Besides, the gauge-fixing just mentioned yields the following equation 
for $x^{-\mu}_A$
\be
(\partial^{++}_w - v^{++}) x^{-\mu}_A = x^{+\mu}_A~. \label{x-eq}
\ee 
These equations are written in the $\tau $-basis $\{x, w \}$ with 
taking account of the definition \p{xbridge}. 

Further, the commutation relation \p{DDcom} in the 
$\lambda $- basis implies the {\it analyticity} of $H^{+3\mu}$, $H^{+4}$, 
i.e. their independence of $x^{-\mu}_A$:
\bea
[(D^{++})_\lambda, ({\cal D}_{\alpha}^+)_\lambda] &=& 0 \quad 
\Rightarrow \nn 
\partial^+_\mu H^{+3\mu} = \partial^+_\mu H^{+4} &=& 0 \quad 
\Rightarrow H^{+3\mu} = H^{+3\mu}(x^+_A, w_A)~, \; 
H^{+4} = H^{+4}(x^+_A, w_A)~, \label{Hanalyt}
\eea
where $({\cal D}_{\alpha}^+)_\lambda $ was defined in \p{Danal}.

The analytic harmonic vielbein $H^{+4}$ is basically just the unconstrained 
QK potential while $H^{+3\mu}$ is expressed in terms of it. To be more 
precise, the QK potential ${\cal L}^{+4}$, as it was defined in \cite{gio}, 
is related to $H^{+4}$ as 
\be
H^{+4} (x^+_A, w_A)= R{\cal L}^{+4}(x^+_A, w_A) + 
Rx^+_\mu H^{+3\mu}(x^+_A, w_A) \label{QKpot} 
\ee
and 
\bea
&& H^{+3\nu} = 
{1\over 2}\Omega^{\nu \mu}\hat{\partial}^-_\mu {\cal L}^{+4}~, 
\label{H3L4} \\
&& \hat{\partial}^-_\mu \equiv \partial^-_\mu + Rx^+_\mu \partial^{--}_A~, 
\label{hatdef}                                        
\eea
where we have explicitly singled out the constant $R$ as 
a contraction parameter to the corresponding HK manifold. Note that 
eqs. \p{QKpot}, \p{H3L4} were 
derived in \cite{gio} after several gauge-fixings of the $\lambda $-world 
gauge freedom, including the 
one soldering analytic diffeomorphisms in $\{x^+_A, w_A \}$ with the 
tangent space analytic $Sp(n)$ rotations. Henceforth, we deal with the QK 
geometry relations in such a maximally gauge-fixed 
framework, with all the pure gauge quantities gauged away.

It can be shown that the only constraint to be satisfied by ${\cal L}^{+4}$ 
is its analyticity, so this object encodes all the information about 
the relevant QK metric, whence its name `QK potential'. Choosing one or 
another explicit ${\cal L}^{+4}$, 
and substituting \p{QKpot}, \p{H3L4} into eqs. \p{H3} - \p{x-eq}, one can 
solve the latter for $x^{\pm \mu}$ and $v^{++}$ as functions of harmonics 
and the $\tau $-basis coordinates $x^{\mu n}$ which now can be interpreted 
as the integration constants of eq. \p{H3} considered as a differential 
equation on the sphere $S^2 \sim w^{\pm i}$ (in each specific case solving 
such equations could be a highly non-trivial task). Having the 
explicit form of the variable change \p{zbridge} - \p{xbridge}, 
it remains to find the appropriate expression of the $\lambda $-world 
vielbeins in terms of ${\cal L}^{+4}$ in order to be able to compute the 
$\tau $-world vielbein and hence the QK metric itself. 

To this end, one should firstly find the $\lambda $-basis form of the 
covariant derivative $D^{--}$ (it equals to $\partial^{--}_u$ in the 
$\tau $-basis $\{ x, v, u \}$). Once again, up to the $z$-derivative 
and $Sp(n)$ connection terms, it reads \cite{gio}
\be
 (D^{--})_\lambda = (z^0)^{-2} \left(\psi \Delta^{--} + \dots \right)~, 
\label{Dminusl} 
\ee
\bea
\Delta^{--} &=& \partial^{--}_A + H^{--+\mu}\partial^-_\mu + 
H^{-3\mu}\partial^+_\mu = 
\frac{1}{1-\partial_w^{--}v^{++}}\;\partial^{--}_w~, \label{Hminus} \\
H^{--+\mu} &=& \Delta^{--}x^{+\mu}_A =
\frac{1}{1-\partial_w^{--}v^{++}}\;\partial^{--}_w x^{+\mu}_A ~, 
\nn  
H^{-3\mu} &=& \Delta^{--}x^{-\mu}_A =
\frac{1}{1-\partial_w^{--}v^{++}}\;\partial^{--}_w x^{-\mu}_A ~, 
\label{defHmin} \\
\psi &=& t^2\left(1-\partial_w^{--}v^{++} \right) = 
{1\over 1 - R\;x^+_\mu H^{--+\mu}} \equiv {1\over 1 - R(x\cdot H)}~.       
\label{defpsi}
\eea
After that one defines the $\lambda $-world derivative 
$({\cal D}^-_\alpha)_\lambda $ by 
\be
({\cal D}^-_\alpha)_\lambda = 
[(D^{--})_\lambda, ({\cal D}^+_\alpha)_\lambda] = (z^{0}_A)^{-1} 
\left(E^{-\mu-}_\alpha(x,w) \partial^+_\mu + 
E^{-\mu+}_\alpha (x,w) \hat{\partial}^-_\mu + \dots \right)~, 
\label{Dxminus}
\ee
where the explicit form of the vielbeins on the r.h.s. can be found in 
\cite{gio}. Let us emphasize that \p{Dxminus} is simply the 
$\lambda $-basis 
form of the $\tau $-basis relation  
\be
[D^{--},  {\cal D}^+_\alpha] = {\cal D}^-_\alpha = 
(z^0)^{-1} \left( w^{-i}e^{\mu n}_{\alpha i}\partial_{\mu n} + ...\right)~. 
\label{Dxmintau}
\ee

Making use of the basic postulate of Riemann geometry, that 
is requiring the commutator 
$[({\cal D}^-_\alpha)_\lambda , ({\cal D}^+_\alpha)_\lambda]$ to contain 
no torsion in $x$-directions, one expresses (after appropriate 
gauge-fixings) 
the $\lambda $-world vielbeins $E^\mu_\alpha $, $E^{-\mu\pm}_\alpha$ in 
\p{Danal}, \p{Dxminus} in terms of the harmonic vielbeins $H^{--+\mu}$, 
$H^{-3\mu}$ and, by eqs. \p{defHmin}, in terms of $v^{++}$, 
$x^{\pm \mu}_A$. Instead of explicitly giving these rather complicated 
expressions, we give here the expressions for the $\lambda $-world metric 
components which are comparatively simple 
\bea
g^{MN}_{(\lambda)} &=& E^{+M\alpha}E^{-N}_\alpha -
E^{-M\alpha}E^{+N}_\alpha~, \qquad M,N = (\nu+, \nu-) \label{defgl} \\
g^{\mu+\;\nu+}_{(\lambda)} &=& 0~, \quad 
g^{\mu+\;\nu-}_{(\lambda)} = g^{\nu-\;\mu+}_{(\lambda)} = 
- E^{-\mu+\alpha}E^{+\nu-}_\alpha  = 
\Omega^{\mu\rho}(\partial\hat H)^{-1\;\nu}_{~~\;\;\rho}~, \nn
g^{\mu-\;\nu-}_{(\lambda)} &=& 
E^{+\mu-\alpha}E^{-\nu-}_\alpha - E^{-\mu-\alpha}E^{+\nu-}_\alpha = 
- 2\;\Omega^{\rho\sigma} (\partial\hat H)^{-1\;\omega}_{~~\;\;\sigma}
(\partial\hat H)^{-1\;(\mu}_{~~\;\;\rho} 
\partial_\omega^+ \hat H^{-3\nu)}~. \label{compmetr}
\eea
Here 
\bea
(\partial \hat H)^{\;\mu}_\nu &\equiv & \partial^+_\nu \hat H^{--+\mu}~, 
\label{def1} \\
\hat H^{--+\mu} &\equiv & {1\over 1 - R(x\cdot H)}\;H^{--+\mu}~, \qquad  
\hat H^{-3\mu} \equiv  {1\over 1 - R(x\cdot H)}\; H^{-3\mu}~. \label{def2}
\eea

Now, making the inverse change of variables $x^{\pm \mu}_A, w^{\pm i}_A 
\rightarrow x^{\mu m}$, $w^{\pm i}$ in $({\cal D}^{\pm}_\alpha)_\lambda $
and comparing the result with \p{Dstruct}, \p{Dxmintau}, \p{metr}, 
taking account of the completeness relation 
$$
w^{+\;i} w^{-\;k} - w^{+\;k}w^{-\;i} = \epsilon^{ki}~,
$$ 
one reads off the expression for the $\tau $-world metric. It is 
as follows 
\be
g^{\mu s \; \nu m} = g^{\omega - \;\sigma -}_{(\lambda)}
\partial^+_\omega x^{\mu s} \partial^+_\sigma x^{\nu m} 
+ g^{\omega + \;\sigma -}_{(\lambda)}\left(
\hat\partial^-_\omega x^{\mu s} \partial^+_\sigma x^{\nu m}
+ \hat\partial^-_\omega x^{\nu m} \partial^+_\sigma x^{\mu s} \right)~,
\label{taumetr}
\ee
where $\hat\partial^-_\mu $ was defined in \p{hatdef}.
All the quantities entering \p{compmetr}, \p{taumetr} can be expressed 
through the $\tau $-basis derivatives of $x^{\pm \mu}_A$ using \p{defHmin}, 
\p{def1}, \p{def2} and the following relations 
\bea
&& \partial^+_\mu = (\partial^+_\mu x^{\rho n}) \nabla_{\rho n}~, \quad 
\nabla_{\rho m} = \partial_{\rho m} + {1\over 1-\partial^{--}_w v^{++}} 
\;(\partial_{\rho m} v^{++}) \partial^{--}_w \label{def3} \\
&& \partial^{\pm}_\mu x^{\rho m}\;\nabla_{\rho m} x^{\pm \nu} = 0~, \qquad 
\partial^{\mp}_\mu x^{\rho m}\;\nabla_{\rho m} x^{\pm \nu} = 
\delta^\nu_\mu~,
\label{matreqs} \\
&& \partial^{--}_A x^{\rho m} = -H^{--+\mu}\partial^-_\mu x^{\rho m} -
H^{-3\mu}\partial^+_\mu x^{\rho m}~. \label{def4}
\eea
Note the useful equation 
\be \label{d++X}
(\partial^{++}_w + v^{++})\partial^+_\mu x^{\rho k} = 0~, 
\ee
which follows from the analyticity-preserving relation  
$[D^{++}, \partial^+_\mu] = 0$ and the obvious property 
$D^{++}x^{\rho k} = 0$. 

In the case of 4-dimensional QK manifolds we will deal with in the sequel
$(\mu, \nu = 1,2 )$, the expression for the $\tau $-basis metric 
\p{taumetr}  can be essentially simplified. After some algebra 
one gets 
\bea
g^{\mu s \; \nu m} &=& {1\over \mbox{det}(\partial \hat{H})}  
\;{1\over [1- R(x\cdot H)](1- \partial^{--}\tilde{v}^{++})} 
\;G^{\mu s \; \nu m}\;, \label{12} \\
G^{\mu s \; \nu m} &=& \epsilon^{\lambda \rho}\;[\;\partial^{--}_w 
X^{+ \mu s}_\lambda \;X^{+ \nu m}_\rho + 
(\mu s \leftrightarrow \nu m)\;] \;, 
\label{13}
\eea
where 
\be 
X^{+ \mu m}_\rho \equiv \partial^+_\rho x^{\mu m} \label{defX}
\ee
are solutions of the system of algebraic equations 
\be \label{1}
X^{+ \mu m}_\rho \nabla_{\mu m} x^{-\nu} = \delta^\nu_\rho, \qquad 
X^{+ \mu s}_\rho \nabla_{\mu s} x^{+\nu} = 0.
\ee   
Note also the useful formula 
\be 
\mbox{det}(\partial \hat{H})  = {1\over [1- R(x\cdot H)]^3} 
\mbox{det}(\partial {H}) \label{det}
\ee
which follows from the relation 
\be 
(\partial \hat H)^\mu_\nu = (\partial H)_\nu^\rho 
\left( \delta^\mu_\rho + R {1\over 1- R(x\cdot H)}\;x^+_\rho H^{--+\mu} 
\right)~. 
\ee

Before closing this Section, we summarize the steps leading from 
the given QK potential ${\cal L}^{+4}$, eq. \p{QKpot}, to the QK metric 
$g^{\mu s\;\nu m}$, eqs. \p{compmetr}, \p{taumetr}, \p{12}, \p{13}. \\

\noindent {\bf A}. Using eqs. \p{QKpot}, \p{H3L4}, one expresses 
the harmonic 
vielbeins $H^{+4}$, $H^{+3\mu}$ through ${\cal L}^{+4}$ and substitute 
these expressions into the differential equations for bridges \p{H3} - 
\p{x-eq}. \\

\noindent {\bf B}. One solves \p{H3} - \p{x-eq} for $v^{++}$, $x^{\pm \mu}$
and obtains the latter as functions of the $\tau $-basis coordinates 
$x^{\mu m}, w^{\pm i}$ (this is the most difficult step). \\

\noindent {\bf C}. One computes the vielbeins $H^{--+\mu}$, $H^{-3\mu} $
using the definition \p{defHmin} and convert their $\partial^+_\mu $ 
-derivatives into $\partial_{\mu m}$ by making use of eq. \p{def3}. \\

\noindent {\bf D}. One computes $\nabla_{\mu n}x^{\pm \rho} $ and finds 
the components of the inverse matrix $\partial^+_\mu x^{\rho m}$, 
$\hat \partial^-_\mu x^{\rho m}$ by solving the algebraic eqs. \p{matreqs} 
or \p{1}. \\

\noindent {\bf E}. One substitutes all that into eqs. \p{compmetr}, 
\p{taumetr}
or \p{12}, \p{13} and finds the explicit form of the QK metric as a 
function of the original coordinates $x^{\mu m}$.

We should stress that these steps go along the same line as in the HK case 
\cite{HK}. The differences are, first, the presence of an extra bridge 
$v^{++}$ and the necessity to solve the corresponding bridge equation 
\p{H4} and, second, the presence of a non-zero constant $R$ in all 
the relations and equations that essentially complicates the computations 
as compared to the HK case. Nevertheless, in both cases, the QK 
and HK ones, 
the basic geometric object is ${\cal L}^{+4}(x^+_A, w_A)$. This implies 
that 
any HK metric has its QK counterparts, the former following from 
the latter via contraction $R\rightarrow 0$. In particular, the flat 
$4n$-dimensional HK manifold has as its QK analogs the compact or 
non-compact homogeneous spaces $Sp(n+1)/Sp(n)\times Sp(1)$ or 
$Sp(n,1)/Sp(n)\times Sp(1)$, depending on the sign of $R$. This case 
corresponds to ${\cal L}^{+4} = 0$ \cite{gio}. In the next Section we 
consider a $4$-dimensional QK manifold with a simplest non-trivial 
${\cal L}^{+4}$, the QK analog of the well-known Taub-NUT 
space \cite{egh}.

\setcounter{equation}{0}
\section{Quaternionic Taub-NUT: bridges and harmonic \break 
vielbeins}
The QK counterpart of the Taub-NUT manifold 
is characterized by the same ${\cal L}^{+4}$ \cite{bgio}
\be 
{\cal L}^{+4} = \left(c_{(\mu\nu)}x^{+\mu}_Ax^{+\nu}_A \right)^2 \equiv 
\left(\phi^{++}\right)^2~. \label{LTN}
\ee
Here $\mu, \nu = (1,2)$ and $c_{\mu\nu} = c_{\nu\mu}$ is a constant 
$3$-vector satisfying the reality condition 
\footnote{We adopt the convention 
$\epsilon_{12} = - \epsilon^{12} = 1$.} 
\be
\overline{(c_{\mu\nu})} \equiv \bar c^{\mu\nu} = \epsilon^{\mu\rho}
\epsilon^{\nu\sigma}c_{\rho\sigma}~. \label{reality}
\ee
The conjugation rules for the coordinates $x^{\pm\mu}$ are as follows 
\footnote{When applied to the harmonic-dependent objects, the complex 
conjugation is always understood as a generalized one, i.e. the 
product of the ordinary conjugation and Weyl reflection of harmonics 
\cite{gikos}.} 
\be
\overline{(x^{\pm \mu})} = - \epsilon_{\mu\nu}x^{\pm \nu} \quad 
\rightarrow \quad 
\overline{\phi^{++}} = \phi^{++} ~.
\ee
They agree with the following reality condition for the $\tau $-basis 
coordinates 
\be
\overline{x^{\mu i}} = \epsilon_{\mu\nu}\epsilon_{ik}x^{\nu k}~.
\ee
In what follows we will frequently use the definition 
\be \label{complx}
x^{+} \equiv x^{+1}~,\;\; \bar x^{+} = \overline{(x^+)} = - x^{+2}~, \;\; 
\overline{(\bar x^+)} = - x^+~.
\ee

Making use of the freedom of constant $SU(2)$ rotations of $x^{+\mu}$ with 
respect to the doublet index $\mu $, one can bring $c_{\mu\nu}$ to 
the form with one non-zero component 
\be  \label{su2frame}
c_{\mu\nu} \rightarrow c_{12} \equiv -i \lambda~, \bar \lambda = \lambda~. 
\ee
In this frame 
\be 
\phi^{++} = -2 i\lambda x^{+1}x^{+2} = 2i \lambda x^+ \bar x^+~. 
\label{phiframe}
\ee
The remnant of the above $SU(2)$ is the $U(1)$ symmetry 
\be
x^{+\;}{}' = e^{i\alpha}\;x^+~, \qquad  \bar x^{+\;}{}' = 
e^{-i\alpha}\;\bar x^+~,
\label{PG}
\ee
which is an obvious invariance of $\phi^{++}$ and ${\cal L}^{+4}$, and 
so it is an isometry of the corresponding QK metric 
(like in the TN case \cite{cmp}). It appeared first in a $N=2$ 
supersymmetry context \cite{{gio},{cmp}} where it was called the 
Pauli-G\"ursey symmetry. This symmetry, like in the TN case \cite{cmp}, 
will essentially help in deducing the QK metric within the present approach. 

To construct the QK metric related to ${\cal L}^{+4}$ \p{LTN} we will 
follow 
the steps listed at the end of the preceding Section. We will omit the 
subscripts `$A$' of the analytic subspace coordinates and the subscript
$`w'$ of the $\tau $-basis partial harmonic derivatives.

At the step {\bf A} we have 
\be
H^{+3\mu} = {1\over 2}\;\epsilon^{\mu\nu}\partial^-_\nu {\cal L}^{+4} = 
2c^\mu_\rho x^{+\rho} \phi^{++}~, \;\; H^{+4} = R\left( {\cal L}^{+4} + 
x^+_\nu H^{+3\nu} \right) = - R(\phi^{++})^2~, \label{HQTN}
\ee
\bea
&& \partial^{++} v^{++} + R(v^{++})^2 = (\phi^{++})^2~,  \label{veq} \\
&& (\partial^{++}  + Rv^{++}) x^{+\mu} = 
2 c^\mu_\rho x^{+\rho} \phi^{++}~. \label{xpluseq} \\
&& (\partial^{++}  - Rv^{++}) x^{-\mu} = x^{+\mu}~. \label{xmineq} 
\eea          
In order to have a well-defined HK limit, we have rescaled $v^{++}$ 
as 
$$
v^{++} \Rightarrow  Rv^{++}~. 
$$

According to the step {\bf B} we should now solve the system of differential 
equations \p{veq} - \p{xmineq}. We will start with \p{veq}, \p{xpluseq}. 
Defining 
\be
v^{++} = \partial^{++}v~, \quad \omega \equiv e^{Rv}~, \quad \hat x^{+\mu} 
\equiv \omega \;x^{+\mu}~, \quad \hat \phi^{++} = c_{\mu\nu} 
\hat x^{+\mu}\hat x^{+\nu} = \omega^2 \;\phi^{++}~, \label{defhat}
\ee 
we rewrite \p{veq}, \p{xpluseq} as 
\bea
&& (\partial^{++})^2 \omega = R\; {(\hat\phi^{++})^2 \over \omega^3} 
\label{omeq}~,  \\
&& \partial^{++} \hat x^{+\mu} = 2 {\hat \phi^{++} \over \omega^2}\;
c^\mu_\rho \hat x^{+\rho} \equiv 2 \kappa^{++}\;
c^\mu_\rho \hat x^{+\rho}~. \label{hatxeq}
\eea
{}From eq. \p{hatxeq} and the definition of $\hat \phi^{++}$ one 
immediately finds 
\be
\partial^{++} \hat \phi^{++} = 0 \quad \Rightarrow \quad \hat \phi^{++} = 
\hat \phi^{ik}(x) w^{+}_iw^+_k~. \label{hatphi}
\ee
We observe that eq. \p{omeq} is none other than the pure 
harmonic part of the equation defining the Eguchi-Hanson metric 
in the harmonic superspace approach \cite{giot}! 
Its general solution was given in \cite{giot}, it depends on four 
arbitrary integration constants, that is, in our case, on four arbitrary 
functions of $x^{\mu i}$. However, these harmonic constants turn out to be 
unessential due to four hidden gauge symmetries of the set of equations 
\p{omeq} - \p{xmineq}. One of them is the scale invariance 
\be
v\;' = v + \beta(x)\;,\;\; \omega\;' = 
e^{R\beta}\omega\;,\;\; \hat \phi^{++\;}{ }' = 
e^{2R\beta} \hat \phi^{++}\;,
\;\; 
v^{++\;}{ }' = v^{++}\;,\; \; x^{\pm \mu\;}{ }' = x^{\pm \mu}\;. 
\label{scale} 
\ee
It is a trivial consequence of the definition of $v$ in eq. \p{defhat}. 
Another symmetry is less trivial, this is a hidden gauge $SU(2)$ symmetry
\bea
&& \delta v = -{1\over R}\alpha^{+-} + \alpha^{--}\partial^{++} v~, \quad 
\delta v^{++} = -{1\over R}\alpha^{++} + \alpha^{--}\partial^{++} v^{++} 
+ 2 \alpha^{+-} v^{++}~, \label{vtransf}  \\
&& \delta x^{\pm \mu} = \alpha^{--}\partial^{++} x^{\pm \mu} \pm \alpha^{+-} 
x^{\pm \mu}~, \label{xtransf}
\eea 
with 
$$
\alpha^{\pm\pm} = \alpha^{(ik)}(x)w^\pm_iw^\pm_k~, \qquad 
\alpha^{+-} = \alpha^{(ik)}(x)w^+_iw^-_k~. 
$$
Using this gauge freedom  one can gauge away four integration constants 
in $\omega $ and write a general solution of eq. \p{omeq} in a fixed gauge 
in the following simple form  
\bea 
&& \omega = \sqrt{1 + R \hat\phi^2} \quad \Rightarrow \quad 
v = {1\over 2R}\; \mbox{ln} ( 1 + R \hat \phi^2 )~,  \label{vslut} \\
&& v^{++} = \partial^{++}v = {\hat \phi\; \hat \phi^{++} \over 
1 +R \hat \phi^2}~, \qquad \hat \phi \equiv
\hat \phi^{(ik)}(x)w^+_i w^-_k~. \label{vplplsol}
\eea 
Of course, one can always restore the general form of the solution as it is 
given in \cite{giot}, 
acting on \p{vslut} by the group \p{scale}, \p{vtransf}, \p{xtransf} 
(with making use of a finite form of these transformations). Note that 
only the rigid subgroup 
$\alpha^{(ik)}= \mbox{const}$ of \p{vtransf}, \p{xtransf} preserves 
the analytic subspace $\{x^{+\mu}_A, w_A^{\pm i} \}$, so the 
$\tau $-world metric is expected to be invariant only under this subgroup. 
Nevertheless, later on we will see that the whole effect of the full gauge 
$SU(2)$ transformation is reduced to the rotation of the metric 
corresponding to the fixed-gauge solution \p{vplplsol} by some 
harmonic-independent non-singular matrix. Thus in what follows we can stick 
to the solution \p{vslut}, \p{vplplsol}.
 
Note that, defining 
\be  \label{defphi2}
2\lambda^2 s^2 \equiv \hat \phi^{(ik)}\hat \phi_{(ik)}~, 
\ee
and using the completeness relation for harmonics, one gets 
an important relation 
\be \label{scalfac1}
1 - R \partial^{--} v^{++} = \left(1 - R \lambda^2 s^2 \right)\; 
{1 - R \hat \phi^2 \over [1 + R \hat \phi^2 ]^2}~. 
\label{expr1}
\ee

Now we are prepared to solve eqs. \p{xpluseq} or \p{hatxeq}. This can be 
done in a full analogy with the HK Taub-NUT case \cite{cmp},
based essentially upon the PG invariance \p{PG}.
Using \p{vplplsol}, we obtain 
\bea
&& \kappa^{++} = \partial^{++} \kappa~,  \nn 
&&(1)\;R > 0\;, \;\; \kappa \equiv \kappa_{(1)} = {1\over \sqrt{R}} 
\;\mbox{arctan} \;\left(\sqrt{R}\,\hat \phi~\right); \label{1kappa} \\ 
&& (2)\;R < 0\;, \;\; \kappa \equiv \kappa_{(2)} = {1\over \sqrt{|R|}} 
\;\mbox{arctanh}\;\left( \sqrt{|R|}\,\hat \phi~\right). \label{2kappa} 
\eea
For definiteness, in what follows we will choose the solution \p{1kappa}. 
Then, passing to the complex notation \p{complx}, choosing the particular 
$SU(2)$ frame \p{su2frame} and making the redefinition 
\be
\hat x^+ = \exp\{2i\kappa\} \tilde x^+~, \qquad 
\overline{\hat x^+} = \exp\{-2i\kappa\} \bar{\tilde x}^+~, \label{deftild}
\ee
we reduce \p{hatxeq} to 
\bea 
&& \partial^{++} \tilde x^+ = 0 \quad \Rightarrow \nn 
&& \tilde x^+ = x^iw^+_i~, \;\;  \bar{\tilde x}^+ = \bar x_i w^{+i} = 
- \bar x^iw^+_i~, \;\; \hat \phi = -2i\lambda x^{(i}\bar x^{k)} 
w^+_iw^-_k~,\label{solut}
\eea
where, in expressing $\hat \phi $, we essentially made use of the 
PG symmetry \p{PG}. Note that the quantity $s$ defined 
in eq. \p{defphi2} is expressed as follows 
\be 
s = x^i\bar x_i~. 
\label{defs}
\ee

Combining eqs. \p{defhat}, 
\p{vslut}, \p{deftild} and \p{solut} we can now write the expressions for 
$x^+$, $\bar x^+$ in the following form 
\be
x^+ = {1\over \sqrt{1 +R \hat \phi^2}} \exp\{2i\kappa\}\; x^iw^+_i~, 
\quad \bar x^+ = - {1\over \sqrt{1 +R \hat \phi^2}} 
\exp\{-2i\kappa\}\; \bar x^iw^+_i~, \label{solutx}
\ee
where $\kappa $ and $\hat \phi $ are expressed through $x^i, \bar x^i$ 
according to eqs. \p{1kappa}, \p{solut}. Comparing \p{solutx} with the 
general definition of the $x$ -bridges \p{xbridge}, we can identify 
$x^i, \bar x^i$ with the $\tau $- world coordinates, i.e. with the 
coordinates of the initial $4$-dimensional QK manifold. 

We still need to find $x^-, \bar x^-$ as functions of $x^i, \bar x^i$ 
and harmonics $w^{\pm}_i$ by solving eq. \p{xmineq}. By means of 
the redefinition 
\be
x^{-\mu} = \sqrt{1 + R \hat \phi^2}\; \hat x^{-\mu} \label{xminredef}
\ee
it is reduced to the form 
\be 
\partial^{++} \hat x^{-} = {1\over 2i\lambda}\;\tilde x^+ 
\frac{\partial}{\partial 
\hat \phi} \exp\{2i\kappa\hat\phi \} 
\label{minhateq}
\ee
(plus a similar equation for $\bar{\hat x}^-$). After this, choosing 
the ansatz
$$
\hat x^- = f(s, \hat \phi)\;\tilde x^-~, \quad  
\bar{\hat x}^- = \bar f\; \bar{\tilde x}^-~,
$$
where 
\be
\tilde x^- = x^iw^-_i~, \qquad \bar{\tilde x}^- = -\bar x^iw^-_i~,
\ee
and exploiting useful identities 
\be 
\hat \phi^{++} \tilde x^- = (\hat \phi + i\lambda s ) 
\tilde x^+~, \quad  
\hat \phi^{--}\tilde x^+ = (\hat \phi  - i\lambda s )
\tilde x^-~, \label{ident}  
\ee
one finds 
\bea
&& \hat x^- = {1\over 2\lambda}\;{1\over (\lambda s) - i\hat \phi }
\left[e^{-2i\kappa(i\lambda s)} - e^
{2i\kappa(\hat \phi)} \right] \tilde x^-~, \nn 
&& \bar{\hat x}^- = {1\over 2\lambda}\;{1\over (\lambda s) + i\hat \phi}
\left[e^{-2i\kappa(i\lambda s)} - e^
{-2i\kappa(\hat \phi)} \right] \bar{\tilde x}^-~. \label{xminsol}
\eea
Here 
\be
\kappa(i\lambda s) \equiv \kappa_0  = 
{i\lambda\over \sqrt{R}}\; \mbox{arctanh}\;\sqrt{R} (\lambda s)~. 
\label{defkap0}
\ee

Now we are ready to fulfill the step {\bf C}, i.e. to find explicit 
expressions for the harmonic vielbein components $H^{--+\mu}, H^{-3\mu}$. 
It is not too illuminating to give here these expressions: given 
the expressions for $x^+$ and $x^-$, eqs. \p{solutx}, \p{xminsol}, 
these objects can be straightforwardly computed using the definition 
\p{defHmin}. Actually, in order to compute the $\tau $-basis metric 
in the $4$-dimensional case, we need only $H^{--+\mu}$, because just 
this quantity enters the expression \p{12}. Moreover, as we will see, 
it appears only inside the scalar factor \p{defpsi}. The latter turns 
out to be surprisingly simple 
\be 
\psi = {1\over 1-R(x\cdot H)} = \left(\frac{1-R\lambda^2 s^2}{1 + Rs + 
R \lambda^2s^2} \right)\frac{1-R \hat \phi^2}{1+ R \hat \phi^2}~.
\label{exprpsi}
\ee 

\setcounter{equation}{0}
\section{Quaternionic Taub-NUT metric}      
At the step {\bf D} we should calculate 
the transition matrix elements $\nabla_{\rho k}x^{\pm}, 
\nabla_{\rho k}\bar{x}^{\pm} $. The computation is straightforward, 
though tiresome. To simplify the formulas, it is convenient to define 
\bea
A(s) \equiv 1 - R\lambda^2s^2~, \quad B(s) \equiv 1+ 
\lambda^2 s (4 + Rs)~, \quad
C(s) \equiv 1+ Rs +R\lambda^2s^2~. \label{defABC}
\eea
Then the results of computation are as follows 
\bea 
\nabla_{\rho k}x^{+} &=& e^{2i\kappa}{1\over 
\sqrt{1+R\hat \phi^2}} \left\{ \delta^1_\rho \left(w^+_k F + 
w^-_k F^{++} \right) 
+ 
\delta^2_\rho \left(w^+_k G + w^-_k G^{++} \right) \right\}~, \label{A} \\
F &=& {1\over 4 A (1-R\hat\phi^2)}\left\{ 
3 A + B + 2i\lambda(3C-A)\hat \phi
+ 2iR\lambda (A +C)\hat \phi^2(\hat \phi + i\lambda s) \right\}\;, 
\label{B} \\
F^{++} &=& -\lambda^2 \left(1 + {C\over A}\right)
(\tilde x^+\tilde{\bar x}^+)~, \quad 
G^{++} = \lambda^2 \left(1+  {C\over A} \right)(\tilde x^+ 
)^2~, \label{C} \\
G &=& -{i\lambda \over A(1-R\hat\phi^2)}\left\{ 
i\lambda(A +C)(1+R\hat \phi^2) + R(A+B)\hat \phi\right\} 
(\tilde{x}^-\tilde{x}^+)~, \label{D} \\
\nabla_{\rho k}x^{-} &=& - {1\over 2i \lambda}{\sqrt{1+R\hat \phi^2}\over 
\hat\phi +i \lambda s} \left\{ \delta^1_\rho \left(w^-_k T + w^+_k T^{--} 
\right) + 
\delta^2_\rho \left(w^-_k S + w^+_k S^{--} \right) \right\}~, \label{AA} \\
T &=& {\lambda^2 \over A}\left\{(4-A-C)e^{-2i\kappa_0} + (A+C)
e^{2i\kappa} \right\}(\tilde x^-\tilde{\bar x}^+)~, \label{BA} \\
T^{--} &=& {i\lambda\over A(1-R\hat\phi^2)} \left\{R(A+B)\hat\phi 
e^{2i\kappa} - 2R(1+2i\lambda\hat\phi)\hat\phi 
e^{-2i\kappa_0} \right. \nn 
&& \left. \qquad  \qquad \qquad    
+\; i\lambda (A+C)(1+R\hat\phi^2)
\left(e^{-2i\kappa_0} - e^{2i\kappa}\right)\right\}
(\tilde x^-\tilde{\bar x}^-)~, 
\label{CA} \\
S &=& \lambda^2\left(1 + {C\over A}\right)
\left(e^{-2i\kappa_0} - e^{2i\kappa}\right)
(\tilde x^+\tilde{x}^-)~, \label{DA} \\
S^{--} &=& {i\lambda\over 2 A(1-R\hat\phi^2)}\left\{\left[4 +(A+B)(A-1) 
-6i\lambda (A+C-2)\hat\phi + 2R(A-1)\hat\phi^2 \right. \right. \nn
&& \left.\left. \qquad \quad  - 2i R \lambda (A+C)\hat\phi^3 
\right] 
\left(\frac{e^{-2i\kappa_0} - e^{2i\kappa}}
{\hat\phi + i\lambda s} \right) 
+8i\lambda (A-1)e^{-2i\kappa_0}\right. \nn
&& \left. \qquad \quad + \;4i\lambda 
\left[2 - A-i\lambda (A+C-2)\hat\phi\right]
\left( e^{-2i\kappa_0} + e^{2i\kappa} \right) \right\} 
(\tilde x^-)^2\;. \label{EA}
\eea
The conjugate quantities $\nabla_{\rho k}\bar x^{\pm}$ follow from the 
above expressions by generalized conjugation. 

Now we should find the entries of the inverse matrix $X^{+\mu i}_\nu 
\equiv 
\partial^+_\nu x^{\mu i}$ by solving the set of algebraic equations 
\p{1}. In the complex notation, this set is divided into the two mutually 
conjugated ones, each consisting of four equations. It is clearly enough 
to consider one such set, e.g. 
\be \label{2}
X^{+ \rho k} \nabla_{\rho k} x^{-} = 1, \quad 
X^{+ \rho k} \nabla_{\rho k} x^{+} = 0, \quad
X^{+ \rho k} \nabla_{\rho k} \bar x^{\pm} = 0~,
\ee
where $X^{+ \rho k} \equiv X^{+ \rho k}_1, 
(\bar X^{+ \rho k} \equiv - X^{+ \rho k}_2) $.
Let us define 
\be \label{3}
\hat X^{+ \rho k} = 
e^{R v} X^{+ \rho k}  = \sqrt{1 + R\hat \phi^2} \;X^{+ \rho k}, 
\quad  (\hat \phi \equiv 
\hat \phi^{+-})~. 
\ee
It satisfies 
\be \label{4}
\partial^{++} \hat X^{+ \rho k} = 0 
\ee
as a consequence of eq. \p{d++X}.

Let us also define 
\be \label{5}
\hat X^{(1)} = \hat X^{+ 1l}w^-_l, \; \hat X^{++(1)} = 
\hat X^{+ 1l}w^+_l, \; 
\hat X^{(2)} = \hat X^{+ 2l}w^-_l, \; \hat X^{++(2)} = 
\hat X^{+ 2l}w^+_l\;.
\ee
Eqs. \p{4} together with the PG $U(1)$ invariance fix these 
quantities up to 
three integration constants which are functions of $s \equiv x^i\bar x_i$: 
\be \label{6}
\hat X^{(1)} = {1\over 2i\lambda}( f_1 + f_3\; \hat \phi ), \; 
\hat X^{++(1)} = (\tilde x^+ \tilde{\bar x}^+)\; f_3, \; 
\hat X^{(2)} = (\tilde{\bar x}^+ \tilde{\bar x}^-)\; f_4, \; 
\hat X^{++(2)} = (\tilde{\bar x}^+)^2 \; f_4\;.     
\ee
Now eqs. \p{2} serve to compute these harmonic constants. After 
substitution 
of the explicit expressions for $\nabla_{\rho k}x^{\pm \nu}$ 
\p{A} - \p{EA}, one finds that the last three eqs. from the set \p{2} 
require 
\be \label{7}
(a)\; f_3 = f_4;\quad (b) \; f_3 = -2i\lambda {A+C \over 3A +B}\; f_1,  
\ee
while the first one uniquely fixes $f_1$
\be \label{8}
f_1 = {i\over 2} \lambda (3A +B)\;e^{2i\kappa_0}~. 
\ee
As a result one gets fairly simple expressions for $\hat X^{(1)}, 
\hat X^{(2)}$
\bea 
\hat X^{(1)} &=& {1\over 4}\{(3A + B) - i\lambda (A+C) \hat \phi \} 
e^{2i\kappa_0}, \; \nn
\hat X^{++(1)} &=& \lambda^2\;(A+C)e^{2i\kappa_0} 
(\tilde x^+ \tilde{\bar x}^+) 
 \;, \nn 
\hat X^{(2)} &=& \lambda^2 \; (A + C) e^{2i \kappa_0}
(\tilde{\bar x}^+ \tilde{\bar x}^-) \;, \nn 
\hat X^{++(2)} &=& \lambda^2 (A + C) e^{2i\kappa_0}(\tilde{\bar x}^+)^2 \;. 
\label{10}    
\eea
It should be pointed out that the system \p{2} is overdetermined, so 
it provides good self-consistency checks of the correctness of the 
expressions for $\nabla_{\rho k} x^{\pm \mu}$. 

It is easy now to restore $\hat X^{+\rho k}$:
\bea 
\hat X^{+1k} &=& \hat{(\partial^+x^k)} = 
{1\over 4}\;[(3A +B)\epsilon^{kl} - 
4\lambda^2(A+C)x^{(k}\bar x^{l)}\;]\;w_l^+
e^{2i\kappa_0}\;, \nn
\hat{\bar{X}}^{+2k} &=& \hat{(\bar \partial^+\bar x^k)} =    
-{1\over 4}\;[(3A + B)\epsilon^{kl} + 
4\lambda^2(A+C)x^{(k}\bar x^{l)}\;]\;w_l^+
e^{2i\kappa_0}\;, \nn
\hat{\bar{X}}^{+1k} &=& \hat{(\bar \partial^+ x^k)} 
= \lambda^2\;(A+C)\;(x^k x^l) w^+_l e^{2i\kappa_0}\;, \nn
\hat{X}^{+2k} &=& \hat{(\partial^+ \bar x^k)} 
= \lambda^2\;(A+C)\;(\bar x^k \bar x^l) w^+_l e^{2i\kappa_0}\;. 
\label{11}
\eea
Recall that 
\be 
\tilde x^{\pm} = x^kw^{\pm}_k, \quad \tilde{\bar x}^{\pm} = - 
\bar x^kw^{\pm}_k, \quad \overline{(x^k)} = \bar x_k, 
\quad \overline{(\bar x^k)} = 
- x_k\;. 
\ee   

Now we have all the necessary bricks to fulfill the last step {\bf E}: to 
compute $\mbox{det} (\partial \hat H)$ 
and, hence, the whole $\tau $ world metric $g^{\rho i, \lambda k}$, 
eq. \p{taumetr}. 

It will be convenient to rewrite \p{taumetr} through $\hat X$, using 
\bea 
&&{1\over (1- Rx^+_\mu H^{--+\mu})(1-R\partial^{--}\tilde{v}^{++})} = 
{1\over C}(1+R\hat{\phi}^2) \label{14}
\eea
(recall eqs. \p{scalfac1}, \p{exprpsi}). 
Then one has 
\bea 
g^{\rho i, \lambda k} &=& {1\over C \mbox{det}(\partial \hat{H})}  
\;\hat{G}^{\rho i, \lambda k}\;, \label{15} \\
\hat G^{\rho i, \lambda k} &=& 
(1+R\hat\phi^2)\; G^{\rho i, \lambda k} = 
\epsilon^{\omega \beta}\;[\;\partial^{--} 
\hat{X}^{+ \rho i}_\omega\;\hat{X}^{+ \lambda k}_\beta + 
(\rho i \leftrightarrow \lambda k) \;]\;. 
\label{16}
\eea
Next, one uses eq. \p{det} to rewrite $\mbox{det} (\partial \hat H)$ 
as follows
\be 
\mbox{det} (\partial \hat H) = {1\over (1 - Rx^+H^{--+})^3}\;
\mbox{det} (\partial H) = \left( {1-R\hat{\phi}^2\over 
1 + R\hat{\phi}^2}\; {A \over C} 
\right)^3\;\mbox{det} (\partial H)\;. \label{17} 
\ee
After some algebra one can cast $\mbox{det} (\partial H)$ into 
the following convenient form 
\bea 
\mbox{det} (\partial H) &=& -{1\over 2}{1\over (1-R\partial^{--} 
\tilde{v}^{++})^2(1 + R\hat \phi^2)}\;[\;
\epsilon^{\alpha \beta}\partial^{--}\hat{X}^{+\rho k}_\alpha 
\partial^{--}\hat{X}^{+\lambda l}_\beta \;]\times \nn 
&& \times [\;\nabla_{\rho k}x^{+\nu}
\nabla_{\lambda l}x^{+\mu}\epsilon_{\nu\mu}\;]~. \label{18} 
\eea
As a result of rather cumbersome, though straightforward computation 
one eventually gets the simple expression for 
$\mbox{det}(\partial \hat H)$
\be
\mbox{det} (\partial \hat H) = A^2\; {B\over C^3}\; e^{4i\kappa_0} = 
(1-R\lambda^2s^2)^2\; 
{1+2\lambda^2 s + \lambda^2 s(2+sR)\over (1+Rs + R\lambda^2s^2)^3}\; 
e^{4i\kappa_0}\;. 
\label{20}
\ee   
As is seen, the harmonic dependence disappeared in $\mbox{det} 
(\partial \hat H)$, as it should be.

The calculation of this determinant is the longest part of the whole 
story. Once this has been done, the computation of the $\tau $ basis 
metric amounts to the computation of entries of the matrix 
$\hat G^{\rho i, \lambda l}$. This can be done rather 
straightforwardly, and the final answer for the inverse metric is  
\bea 
g^{1k, 1t} &=& -{C^2D\over A^2 B}\;\; (x^kx^t)\;, \label{1k1t} \\ 
g^{2k, 2t} &=& -{C^2D\over A^2B}\;\; (\bar x^k \bar x^t)\;, \label{2k2t} \\
g^{1k, 2t} &=& {C^2\over A^2B}\;[- \epsilon^{kt} A^2 
+ D(x^k\bar x^t)\;]\;, \label{1k2t} 
\eea 
where $D \equiv \lambda^2 (A+C)(A+B) = 2\lambda^2 (2+Rs)(1+2\lambda^2s)$.
 
The metric tensor is then readily obtained
\be\label{dist}
\left\{\begin{array}{ll} 
\displaystyle g_{1k, 1t}\ = & \displaystyle \frac{D}{C^2 B}\; 
(\bar x_k \bar x_t)\;,  \\ [3mm]
\displaystyle g_{2k, 2t}\ = & 
\displaystyle \frac{D}{C^2 B}\; ( x_k x_t)\;,  \\ [3mm]
\displaystyle g_{1k, 2t}\ =& \displaystyle \frac 1{C^2 B}\;
[B^2 \epsilon_{kt} + D(\bar x_k x_t)\;]\;, \end{array}\right. 
\ee
thanks to the identity $A^2+2sD=B^2$. Note that the final expressions 
for the metric and its inverse are valid for any sign of the parameter 
$R$ even if, on the intermediate steps, the choice of sign was 
essential.

As the last topic of this Section we give how the metric \p{taumetr}
is transformed under the gauge $SU(2)$ transformations \p{vtransf}, 
\p{xtransf}. Exploiting only the fact that in the present case 
${\cal L}^{+4}$, $H^{+4}, H^{+3\mu}$ contain no explicit harmonics 
in the analytic basis, one can show that 
\be 
\delta g^{\mu m\; \nu k} = \alpha^{--}\partial^{++} g^{\mu m\; \nu k} 
+ g^{\sigma p\; \nu k} \Lambda_{\sigma p}^{\mu m} (x) + 
g^{\mu m\; \sigma p} \Lambda_{\sigma p}^{\nu k}~, \label{su2trang}
\ee
where 
\be \label{Lambda}
\Lambda_{\sigma p}^{\nu k}(x) = \partial_{\sigma p} \alpha^{(lt)}(x) \; 
\left( T^A_{(lt)}x^{\nu k} \right)~, \quad T^A_{(lt)} \equiv 
w_{A\;(l}^+\frac{\partial}{\partial w^{+ t)}_A} + 
w_{A\;(l}^-\frac{\partial}{\partial w^{- t)}_A}~. 
\ee
The property that $\Lambda_{\sigma p}^{\nu k}$ does not depend on 
harmonics in the $\tau $ -basis, $\partial^{++} \Lambda_{\sigma p}^{\nu k} 
= 0$, is a direct consequence of the relation $[D^{++}, T^A_{(lt)}] = 0$ 
which follows from the fact that the analytic harmonic vielbeins bears 
no dependence on $w^{\pm i}_A$ in the case at hand. The first term in 
\p{su2trang} vanishes because $\partial^{\pm\pm}g^{\mu s \; 
\nu m} = 0$. Thus in the rigid case ($\partial_{\rho m}\alpha^{(ik)} = 0$) 
the metric is not changed at all, while in the general case it 
undergoes some rotation by each of its world indices by a non-singular 
matrix $\delta^{\sigma}_\mu \delta^{p}_k + \Lambda_{\sigma p}^{\nu k}(x) + 
\dots $. Note that, similarly to the HK TN case, the $SU(2)$ factor 
of the $U(2)$ isometry group of the QTN metric 
(it rotates $x^i $ and $\bar x^i$ by their doublet indices) 
in the harmonic space formulation originates from a rigid $SU(2)$ 
that rotates the doublet indices of the harmonics 
$w^{\pm i}$ and $ w_A^{\pm i}$. The analytic coordinates 
$x^{\pm \mu}_A$ and the bridge $v^{++}$ behave under this $SU(2)$ 
as scalars of zero weight.

\setcounter{equation}{0}
\section{Limiting cases and identification with the known metrics}
To compare to the results in the literature \cite{egh} one has to use 
\cite{dev}
\be\label{sig1}
dx^i=x^i\left(\frac{ds}{2s}+i\frac{\sigma_3}{2}\right)-
\bar x^i\left(\frac{\sigma_2-i\sigma_1}{2}\right),\qquad
d\sigma_i=\frac 12\epsilon_{ijk}\,\sigma_j\wedge\, \sigma_k.
\ee
Using the notation $\ A\cdot B\equiv A^i\, B_i\ $ relation (\ref{sig1})
implies 
\be\label{sig2}
-\bar x\cdot dx=\frac{ds}{2}+
is\frac{\sigma_3}{2},\qquad dx\cdot d\bar x=\frac{ds^2}{4s}+
\frac s4(\sigma_1^2+\sigma_2^2+\sigma_3^2),\quad s=x\cdot\bar x.\ee
The metric given by (\ref{dist}) writes
\be\label{dist1}
\frac D{C^2 B}\left[(\bar x\cdot dx)^2+(x\cdot d\bar x)^2\right]
+2\frac B{C^2}dx\cdot d\bar x
+2\frac D{C^2 B}(\bar x\cdot d x)(x\cdot d\bar x),\ee
and becomes
\be\label{dist2}
\frac 12\left[ \frac B{sC^2}\, ds^2+\frac{sB}{C^2}\,
(\sigma_1^2+\sigma_2^2)
+\frac{sA^2}{C^2 B}\,\sigma_3^2\right].\ee

\noindent Three limiting cases give metrics of interest
\footnote{We omit the overall $1/2$ factor.} :
\begin{remunerate}
\item The $\lambda=0$ limit leads to
\be\label{dist3}
\frac 1{(1+Rs)^2}\,\left(\frac{ds^2}{4s}+
\frac s4(\sigma_1^2+\sigma_2^2+\sigma_3^2)\right)=
\frac{dx\cdot d\bar x}{(1+Rs)^2}.\ee
For $R>0$ we have the compact coset $HP^1\sim S^4\sim SO(5)/SO(4)$ 
in its conformally flat form (the so- called  de Sitter metric). For 
$R<0$ we are left with 
the hyperbolic metric on the unit ball i. e. 
the  non-compact symmetric coset $SO(4,1)/SO(4)$ ( the anti-de Sitter
 metric).
\item The $R=0$ limit leads to
\be\label{dist4}
(1+4\lambda^2s)\frac{ds^2}{s}+s(1+4\lambda^2s)\,(\sigma_1^2+\sigma_2^2)
+\frac s{1+4\lambda^2s}\,\sigma_3^2,\ee
which we recognize as the Taub-NUT metric which is complete for $s\geq 0$ 
and $4\lambda^2\geq 0$.
\item The $R=4\lambda^2$ limit leads to
\be\label{dist5}
\frac{ds^2}{s(1+Rs/2)^2}
+\frac s{(1+Rs/2)^2}\,(\sigma_1^2+\sigma_2^2)
+\frac{s(1-Rs/2)^2}{(1+Rs/2)^4}\,\sigma_3^2 ,\ee
which is not complete for $R>0$. For $R<0$ the change of coordinates 
$\displaystyle r^2=\frac{-Rs}{(1-Rs/2)^2}$ brings the metric to the form
\be\label{dist6}
\frac 2{(-R)}\left[\frac{(dr)^2}{(1-r^2)^2}
+\frac{r^2}{1-r^2}\frac{\sigma_1^2+\sigma_2^2}{4}
+\frac{r^2}{(1-r^2)^2}\frac{\sigma_3^2}{4}\right].\ee
If we take $0\leq r<1$ then this is Bergmann's metric in the unit ball of  
${\mathbb C}^2$ which is complete. It does correspond to the non-compact 
K\"{a}hler symmetric space $SU(2,1)/U(2)$.
\end{remunerate}

The most general Bianchi IX euclidean Einstein metrics can be deduced from 
Carter's results \cite{ca}. A convenient  standardization \cite{chv} is the 
following
\be\label{ca1}
d\tau^2=l^2\left \{\frac{r^2-1}{\Delta(r)}(dr)^2+4\  
\frac{\Delta(r)}{r^2-1} \ \sigma_3 ^2 
+(r^2-1) (\sigma_1 ^2 +\sigma_2 ^2)\right \},\ee
with
\be\label{ca2}
\Delta(r)=
\frac{-\Lambda l^2}{3}r^4+(1+2\Lambda l^2)r^2 -2\,M\,r+1+\Lambda l^2.\ee
These metrics are Einstein, with Einstein constant $\Lambda$ and 
isometry group $U(2)$.

In dimension 4 a quaternionic metric is defined to be an Einstein metric 
with self dual Weyl tensor. A simple computation gives
\be\label{weyl}
\displaystyle W^+=\frac {1}{l^2}
\frac{(-1+M-\frac 43\Lambda l^2)}{(r-1)^3}\ Y,\qquad
Y=\left(\begin{array}{ccc}
1 & 0 & 0 \\
0 & 1 & 0 \\
0 & 0 & -2\end{array}\right).\ee

Imposing $\ W^+\ =0$ and getting rid of the parameter $M$ gives for 
the metric
\be\label{ca3}
d\tau^2({\mathbb Q})=l^2\left \{\frac{r+1}{r-1}\frac{(dr)^2}{\Sigma(r)}
+4\  \frac{r-1}{r+1}\,\Sigma(r)\,\sigma_3 ^2
+(r^2-1) (\sigma_1 ^2 +\sigma_2 ^2)\right \},\ee
where now
\be\label{ca4}
\Sigma(r)=1-\frac{\Lambda\, l^2}{3}(r-1)(r+3).\ee
The identifications
\be\label{ca5}
\frac{r-1}{2}=(4\lambda^2-R)\frac s{1+Rs+R\lambda^2s^2},\qquad
\frac 43{\Lambda l^2}=\frac R{4\lambda^2-R},\ee
give the relation
\be\label{ca6}
4(4\lambda^2-R)\left[ \frac B{sC^2}\, ds^2
+\frac{sB}{C^2}\,(\sigma_1^2+\sigma_2^2)
+\frac{sA^2}{C^2 B}\,\sigma_3^2\right]=\frac{d\tau^2({\mathbb Q})}{l^2}.\ee

Note that the quaternionic metric \p{ca3} is complete for $\Lambda < 0$ 
and is asymptotiaclly Anti-de Sitter. It has been recently considered 
in \cite{hhp} under the name Taub-NUT-AdS metric and reveals itself a useful
background for computing black-holes entropy.

For the sake of completeness let us mention that this (Weyl) self-dual 
Einstein metric was also derived in \cite{ped} in the form
$$d\tau^2_1=\frac 1{(1-t^2)^2}\left\{\frac{1+m^2t^2}{1+m^2t^4}(dt)^2
+\frac{t^2(1+m^2t^4)}{1+m^2t^2}\,\frac{\sigma_3^2}{4}
+t^2(1+m^2t^2)\,\frac{(\sigma_1^2+\sigma_2^2)}{4}\right\}.$$
The identifications
$$t^2=\frac{s-1}{s+2m^2+1},\qquad -\frac 43\Lambda l^2=\frac 1{m^2+1},$$
lead to the relation
$$d\tau^2({\mathbb Q})=16(m^2+1)\cdot d\tau^2_1~.$$ 

\setcounter{equation}{0}
\section{Conclusion}
In this paper we used the harmonic space formulation of the 
QK geometry \cite{gio} to compute the QK metric with a 
non-trivial quaternionic potential ${\cal L}^{+4}$, 
the four-dimensional quaternionic Taub-NUT metric. 
We found that the harmonic space techniques, 
like in the HK case \cite{{cmp},{HK},{giot}}, allows one 
to get the {\it explicit} form of the QK metric starting from a given 
QK potential and following a generic set of rules. It would be 
interesting to apply this approach to find the QK analogs of 
some other interesting 4- and higher-dimensional HK metrics, in particular, 
the quaternionic Eguchi-Hanson metric and the quaternionic generalization  
of the multicentre metrics of Gibbons and Hawking \cite{gh}.

A first simple generalization of the Taub-NUT potential would be to 
add the dipolar breaking \cite{gorv}
$$
{\cal L}^{+4} = \eta^{--}(\phi^{++})^3~, \qquad \eta^{--} = \eta^{(ik)}
u^-_iu^-_k~, \;\; \eta^{(ik)} = \mbox{const}~.
$$ 
When trying to compute the quaternionic metric with the same 
${\cal L}^{+4}$
along the lines of Sect. 3, one encounters a difficulty already at the  
step of solving the equation for the relevant bridge $v^{++}$. Making 
the same changes of variables as those leading to eq. \p{omeq}, we get 
in this case 
$$
(\partial^{++})^2 \omega = 
2R \;\eta^{--}\;{(\hat\phi^{++})^3 \over \omega^5}~, 
\quad \partial^{++}\hat\phi^{++} = 0~.
$$
This equation is not so easy to solve as compared to \p{omeq}. Thus, 
to attack the multicenter case, it will perhaps prove advantageous 
to develop some ways around, e.g., the harmonic space version 
of the quaternionic quotient construction \cite{galic}. 

\vspace{0.5cm}

\noindent{\Large\bf Acknowledgements} 
\vspace{0.3cm}

\noindent E.I. thanks the Directorate of LPTHE and the Universit\'e Paris 7, 
for the hospitality extended to him during the course of this work. His 
work was partly supported by the grant of Russian Foundation of Basic 
Research RFBR 96-02-17634 and by INTAS grants INTAS-93-127-ext, 
INTAS-96-0538, INTAS-96-0308.


\begin{thebibliography}{234}
\bibitem{gikos} A. Galperin, E. Ivanov, S. Kalitzin, V. Ogievetsky and 
E. Sokatchev, {\sl Class. Quantum Grav.} {\bf 1} (1984) 469.
\bibitem{cmp} A. Galperin, E. Ivanov, V. Ogievetsky and E. Sokatchev,
{\sl Commun. Math. Phys.} {\bf 103} (1986) 515.
\bibitem{HK} A. Galperin, E. Ivanov, V. Ogievetsky and E. Sokatchev, 
{\sl Ann. Phys.} (N.Y.) {\bf 185} (1988) 22.
\bibitem{gio} A. Galperin, E. Ivanov and O. Ogievetsky, 
{\sl Ann. Phys.} (N.Y.) {\bf 230} (1994) 201.
\bibitem{giot} A. Galperin, E. Ivanov, V. Ogievetsky and P.K. Townsend, 
{\sl Class. Quantum Grav.}, {\bf 7} (1986) 625.
\bibitem{gorv} G.W. Gibbons, D. Olivier, P.J. Ruback and G. Valent, 
{\sl Nucl. Phys.}, {\bf B 296} (1988) 679.
\bibitem{bw} J. Bagger and E. Witten, {\sl Nucl. Phys.} 
{\bf B 222} (1983) 1.
\bibitem{iv1} E. Ivanov, G. Valent, `Quaternionic Taub-NUT from the harmonic 
space approach', Preprint LPTHE 98-43, JINR E2-98-248, September 1998; 
{\tt hep-th/9809108}.
\bibitem{egh} T. Eguchi, B. Gilkey and J. Hanson, {\sl Physics Reports}, 
{\bf 66}, No. 6 (1980) 213.
\bibitem{HK1} A. Galperin, E. Ivanov, V. Ogievetsky and E. Sokatchev,
{\sl Ann. Phys.} (N.Y.) {\bf 185} (1988) 1.
\bibitem{bgio} J.A. Bagger, A.S. Galperin, E.A. Ivanov and 
V.I. Ogievetsky,
{\sl Nucl. Phys.} {\bf B 303} (1988) 522.
\bibitem{dev} F. Delduc and G. Valent, {\sl Class. Quantum Grav.}, {\bf 10} 
(1993) 1201.
\bibitem{ca} B. Carter, {\sl Commun. Math. Phys.} {\bf 10} (1968) 280.
\bibitem{chv} T. Chave and G. Valent, {\sl Class. Quantum Grav.}, {\bf 13}
 (1996) 2097.
\bibitem{ped} H. Pedersen, {\sl Math. Ann.}, {\bf 274} (1986) 35.
\bibitem{hhp} S.W. Hawking, C.J. Hunter and Don N. Page, 'Nut Charge,
Anti-de Sitter Space and Entropy', {\tt hep-th/9809035}.
\bibitem{gh} G. Gibbons and S. W. Hawking, {\sl Phys. Lett.} {\bf B 78} 
(1978) 430.
\bibitem{galic} K. Galicki, {\sl Commun. Math. Phys.} {\bf 108} 
(1987) 117; {\sl Class. Quantum Grav.}, {\bf 8} (1991) 1529. 

\end{thebibliography}
\end{document}